\documentclass[aps,prd,twocolumn,showpacs]{revtex4-1}%
\usepackage{epsfig}
\usepackage{color}
\usepackage{amsmath}
\usepackage{amssymb}
\usepackage{amsfonts}
\usepackage{bm}
\usepackage{dcolumn}
\usepackage{graphicx}
\usepackage{subfigure}
\usepackage{natbib}
\usepackage{mciteplus}
\usepackage{float}
\usepackage[capitalize]{cleveref}%
\providecommand{\U}[1]{\protect\rule{.1in}{.1in}}

\begin{document}
\title{Piezoelectric scattering of optical polaron in halide perovskites}
\author{Ming-Hui Zhang$^{1\dagger}$ and Ming-Liang Zhang$^{2}$}
\affiliation{$^{1}$ College of Mechanical and Electronic Engineering, Shandong University of Science and Technology, Qingdao, 266590, China}
\affiliation{$^{2}$ School of Mechanical Engineering, Tianjin University, Tianjin City, 300072, China}


\begin{abstract}
For the intrinsic carriers of MAPbBr$_{3}$, the temperature $T$ dependent
mobility $\mu(T)$ of behaves like $\mu\propto T^{-1/2}$ in piezoelectric
tetragonal phase, $\mu\propto T^{-1.4}$ in non-piezoelectric cubic phase. But
for the photo-generated carriers in other halide perovskites ABX$_{3}$,
$\mu\propto T^{-3/2}$ behavior is typical. Due to the strong interaction of
carrier with longitudinal optical phonon, in ABX$_{3}$ the carriers mainly
exist as optical polarons. The softness of ABX$_{3}$ renders it without
inversion center in tetragonal phase, which allows piezoelectric effect at low
carrier concentration. The variations of $\mu(T)$ behavior results from (1)
the wave vector dependence of the piezoelectric interaction of polarons with
acoustic phonons is different from that of ordinary polaron-acoustic phonon
interaction; (2) the residual interaction of polaron with 2 longitudinal
optical phonons can be ignored at low temperature, but is important at higher
temperature; and (3) the concentration of intrinsic carriers is determined by
temperature, while the concentration of photo-generated carriers is determined
by the incident flux of photons.

\end{abstract}
\maketitle

\section{introduction}

In the photovoltaic application of halide perovskites ABX$_{3}$ (A=MA$^{+}$,
FA$^{+}$, Cs$^{+}$; B=Pb$^{2+}$, Sn$^{2+}$; X=I$^{-}$, Br$^{-}$, Cl$^{-}$),
one of the key parameters is the temperature $T$ dependent carrier mobility
$\mu(T)$ \cite{don,xing344,stran13}. ABX$_{3}$ is an ionic crystal, due to the
strong interaction of electron (hole) with longitudinal optical (LO) phonons,
electrons (holes) mainly exist as optical electronic (hole) polarons
\cite{pro,fro14,zhu,bre,sen,ost17,mob1}. For the photo-generated carriers in
non-piezoelectric phase, $\mu(T)\propto T^{-3/2}$ is typical
\cite{oga,mil,sav}, which results from: (1) the screening to the interactions
of polaron with other elementary excitations is primarily caused by the
displacements of ions \cite{ono,mob1}; (2) the change in the momentum
distribution function of polaron is dominated by the 1-longitudinal acoustic
(LA) phonons emission or absorption; and (3) the concentration $n_{e}$ of
electrons (holes) is fixed by the incident flux of photons \cite{mob1}.
Recently, for the \textit{intrinsic} carriers in MAPbBr$_{3}$, a different
behavior is observed: $\mu(T)\propto T^{-1/2}$ for tetragonal phase
$160$K$<T<236$K; and $\mu(T)\propto T^{-1.4}$ for cubic phase $236$K$<T<280$K
\cite{yihal}. It is compelling to investigate: (1) the reasons lead to
different $\mu(T)$ behaviors in above two temperature regimes; and (2) why the
transport behavior of \textit{intrinsic} carriers in MAPbBr$_{3}$ is different
to that of the photo-generated carriers in other halide perovskites.

The outline of this work is as following: in Sec.\ref{piepa}, we first
determine the existing condition of piezoelectric effect in MAPbBr$_{3}$. The
appearance of piezoelectric effect in a superparaelectric state makes the
polarization-strain tensor and dielectric function increase with the decrease
of temperature. In Sec.\ref{eap}, we contrast the piezoelectric coupling
$H_{\text{e-pie}}$ of electron with acoustic phonon in piezoelectric phase and
the ordinary electron-acoustic phonon interaction $H_{\text{e-Aph}}$ in
non-piezoelectric phase. In Sec.\ref{eop}, by comparing the strengths of
$H_{\text{e-pie}}$ and $H_{\text{e-Aph}}$ with the strength of electron-LO
phonon interaction $H_{\text{e-LO}}$, we argue that the electrons (holes)
mainly exist as optical polaron in both piezoelectric tetragonal phase and
non-piezoelectric cubic phase. In Sec.\ref{pp}, we determine the dominant
interaction(s) controlling charge transport by comparing the residual 2-LO
phonon interaction $H_{\text{P-LO}}$ of polaron with the piezoelectric
polaron-acoustic phonon interaction $H_{\text{P-pie}}$ in piezoelectric phase,
and with\ the ordinary polaron-LA phonon interaction $H_{\text{P-LA}}$ in
non-piezoelectric phase. In Sec.\ref{mpie}, by analyzing the collision
integral in Boltzmann equation, we find that the $\mu(T)\propto T^{-1/2}$
behavior in tetragonal phase of results from two reasons: (1) each term in
$H_{\text{P-pie}}$ is proportional to $k^{-1/2}$, where $k=|\mathbf{k}|$,
$\mathbf{k}$ is the wave vector of acoustic phonon; and (2) the equilibrium
distribution $f_{0\mathbf{p}}$ of polarons is given by $\exp(-\hbar^{2}%
p^{2}/2m_{\text{P}}k_{B}T)$ \cite{ash}, where $p=|\mathbf{p}|$, $\mathbf{p}$
is wave vector of polaron. In Sec.\ref{mni}, we show that for
\textit{intrinsic} carriers, $H_{\text{P-LA}}$ leads to $\mu(T)\propto T^{-1}%
$, $H_{\text{P-LO}}$ leads to $\mu(T)\propto T^{-3/2}$. Finally, we explain
why $\mu(T)$ for the \textit{intrinsic} carriers is different to $\mu(T)$ for
the photo-generated carriers.

\section{interplay of piezoelectric and superparaelectric}

\label{piepa}

From the phase diagram of MAPbBr$_{3}$, we induct the existing temperature
range of piezoelectricity. Above 236K, MAPbBr$_{3}$ is in cubic phase
\cite{ono}, which cannot be statically piezoelectric due to the existence of
inversion center \cite{v8}. For $149$K$<T<236$K, MAPbBr$_{3}$ is in tetragonal
phase; below 149K, orthorhombic phase \cite{ono}. Because of the softness of
MAPbBr$_{3}$, even in tetragonal phase inversion center cannot exist. Both
tetragonal and orthorhombic phases of MAPbBr$_{3}$ lack of inversion center,
which is a necessary condition to be a piezoelectrics \cite{v8}.

The temperature dependent dielectric function \cite{ono} help us to surmise
the superparaelectricity and ferroelectricity. Denote $\varepsilon(\omega,T)$
as the frequency $\omega$ and temperature $T$ dependent dielectric function.
Above 149K, the observed $\varepsilon(0,T)$ can be described by Curie--Weiss
law, i.e. MAPbBr$_{3}$ is in a superparaelectric phase \cite{ono}; below 149K,
$\varepsilon(0,T<149$K$)\thickapprox\varepsilon_{\infty}$, the optical
dielectric constant caused by bound electrons \cite{ono}. The above facts
indicates that: (1) If MAPbBr$_{3}$ continued staying in tetragonal phase
below 149K rather than become an orthorhombic phase, it would be in a
ferroelectric phase below a critical temperature $T_{c}$, $T_{c}$ is a very
small positive number;\ (2) the rigid orthorhombic phase greatly limited the
displacements of B$^{2+}$ and X$^{-}$ ions, only the bound electrons cause
screening, ferroelectric phase cannot be realized even at $T=0$K. Because
piezoelectric effect requires shifts of ions, it is also impaired in
orthorhombic phase. Summing up above discussion, for$\ 149$K$<T<236$K,
tetragonal MAPbBr$_{3}$ is both piezoelectric and superparaelectric.

When a material is in a state which is both piezoelectric and
superparaelectric, its dielectric function and polarization-strain tensor has
unusual temperature dependence. Denote $\widetilde{\Phi}$ as the thermodynamic
potential density with electric field and stress tensor as independent
variables. Then for $T>T_{c}$, one has \cite{v8}:
\[
\widetilde{\Phi}=\Phi_{0}+A(T-T_{c})P_{z}^{2}/\epsilon_{0}+BP_{z}^{4}-\Lambda
P_{z}\sigma_{xy}%
\]%
\begin{equation}
-\mu^{(P)}\sigma_{xy}^{2}-E_{z}P_{z}-\frac{\epsilon_{0}}{2}E_{z}%
^{2},\label{tpo}%
\end{equation}
where $P_{z}$ is the $z-$component of polarization vector, $E_{z}$ is the
external electric field, $\Lambda$ is the piezoelectric coefficient,
$\sigma_{xy}$ is the $xy-$compoent of stress tensor, $\mu^{(P)}$ is the
modulus for constant polarization $P_{z}$, $A$ and $B$ are usual expansion
coefficients of free energy for paraelectric-ferroelectric transition. Using
$\partial\widetilde{\Phi}/\partial P_{z}=0$ and neglecting $P_{z}^{3}$ term,
the equilibrium polarization is \cite{v8}:%
\begin{equation}
P_{z}=\frac{\epsilon_{0}(E_{z}+\Lambda\sigma_{xy})}{2A(T-T_{c})}.\label{po}%
\end{equation}
In a piezoelectric material, the polarization-strain tensor $\Xi_{\gamma
\alpha\beta}$ is defined by $P_{\gamma}=\Xi_{\gamma\alpha\beta}u_{\alpha\beta
}$, where $u_{\alpha\beta}$ is the strain tensor, $\alpha$, $\beta$,
$\gamma=x$, $y$, $z$. The strain $u_{\alpha\beta}$ of a material specifies the
configurations of atoms, the later directly relates to the polarization
$P_{\gamma}$ caused by the displacements of atoms. Therefore $\Xi
_{\gamma\alpha\beta}$ is an inherent quantity. The polarization-stress tensor
$\Gamma_{\gamma\alpha\beta}$ is defined by $P_{\gamma}=\Gamma_{\gamma
\alpha\beta}\sigma_{\alpha\beta}$, where $\sigma_{\alpha\beta}$ is the stress
tensor \cite{lica74}. $\sigma_{\alpha\beta}$ relates to $u_{\gamma\delta}$ by
$\sigma_{\alpha\beta}=c_{\alpha\beta\gamma\delta}u_{\gamma\delta}$, where
$c_{\alpha\beta\gamma\delta}$ is the elastic constant tensor. Repeated indices
are implicitly summed over. One can easily see the relation between two
piezoelectric coefficients:
\begin{equation}
\Xi_{\gamma\xi\eta}=\Gamma_{\gamma\alpha\beta}c_{\alpha\beta\xi\eta
},\label{eg}%
\end{equation}
or
\begin{equation}
\Gamma_{\gamma\xi\eta}=\Xi_{\gamma\alpha\beta}(c^{-1})_{\alpha\beta\xi\eta
},\label{eg1}%
\end{equation}
where $\xi$, $\eta=x$, $y$, $z$ \cite{pib}. Eq.(\ref{eg1}) tell us that softer
materials like halide perovskites have a larger electrical response to stress.
Later we will suppress the indices of tensor if not cause confusion. From
Eq.(\ref{po}), one has $\Gamma_{zxy}=\epsilon_{0}\Lambda\lbrack2A(T-T_{c}%
)]^{-1}$. By means of Eq.(\ref{eg}), the polarization-strain tensor is
\begin{equation}
\Xi=c\epsilon_{0}\Lambda\lbrack2A(T-T_{c})]^{-1}.\label{e1}%
\end{equation}
Comparing to factor $(T-T_{c})^{-1}$ in Eq.(\ref{e1}), the temperature
dependence of $c_{\alpha\beta\gamma\delta}$ is very weak \cite{var70}, we
shall view $c\Lambda$ as a temperature-independent constant.

For $T>$236K, MAPbBr$_{3}$ is still superparaelectric but no longer
piezoelectric. $\widetilde{\Phi}$ is simplified to \cite{v8}%
\begin{equation}
\widetilde{\Phi}=\Phi_{0}+A(T-T_{c})P_{z}^{2}/\epsilon_{0}+BP_{z}^{4}%
-\frac{\epsilon_{0}}{2}E_{z}^{2}.\label{tpa}%
\end{equation}
Eqs.(\ref{tpo},\ref{tpa}) imply that for both tetragonal phase and cubic
phase,
\begin{equation}
\varepsilon(\omega,T)=\varepsilon_{\infty}+\frac{1}{2A(T-T_{c})}.\label{di1}%
\end{equation}
Eq.(\ref{di1}) could reproduce the observed $\varepsilon(0,T)$\ \cite{ono}.
The form of $\varepsilon(\omega,T)$ is the same for both the piezoelectric
tetragonal phase and non-piezoelectric cubic phase. For a halide perovskite
not in orthorhombic phase \cite{ono}, the screening caused by bound electrons
is much smaller than that caused by the displacements of ions, i.e.
$\varepsilon_{\infty}\ll\lbrack2A(T-T_{c})]^{-1}$, \ one has%
\begin{equation}
\varepsilon(\omega,T)\thickapprox\frac{1}{2A(T-T_{c})}.\label{di2}%
\end{equation}

\section{electron-acoustic phonon interaction}

\label{eap}

The piezoelectric carrier-acoustic phonon interaction is different to the
ordinary carrier-acoustic phonon interaction in a non-piezoelectric phase.
When a material is piezoelectric, the interaction $H_{\text{e-pie}}$ of a
electron (hole) at position $\mathbf{x}$ with the acoustic phonons is
\cite{lica74}%
\begin{equation}
H_{\text{e-pie}}=\sum_{s\mathbf{k}}\frac{V_{\mathbf{k}s}^{\text{pie}}}%
{\sqrt{\mathcal{N}}}(a_{s\mathbf{k}}e^{i\mathbf{k}\cdot\mathbf{x}%
}+a_{s\mathbf{k}}^{\dag}e^{i\mathbf{k}\cdot\mathbf{x}}),\label{epi}%
\end{equation}
where%
\begin{equation}
V_{\mathbf{k}s}^{\text{pie}}=\frac{q\overline{\Xi}_{\gamma}e_{\gamma}%
^{s}(\mathbf{k})}{2\overline{\varepsilon}\epsilon_{0}}\sqrt{\frac{\hbar
}{2Mc_{s}k}},\label{epi1}%
\end{equation}
is the coupling energy of electron (hole) with acoustic phonon $|s\mathbf{k}%
\rangle$, $s=l$, $t_{1}$, $t_{2}$ is the index of longitudinal and two
transverse acoustic branches, $c_{s}$ is the speed of sound of the $s$th
branch. $M$ is the total mass of the atoms in a primitive cell. $e_{\gamma
}^{s}(\mathbf{k})$ is the $\gamma$th component of polarization vector of a
$|s\mathbf{k}\rangle$ phonon, $a_{s\mathbf{k}}$ is the annihilation operator
for $|s\mathbf{k}\rangle$ phonon. Denote $k_{\alpha}$ as $\alpha$th component
of wave vector $\mathbf{k}$. $\overline{\Xi}_{\gamma}=\widehat{k}_{\alpha}%
\Xi_{\gamma\alpha\beta}\widehat{k}_{\beta}$ is the reduced polarization-strain
coefficient, $\widehat{k}_{\alpha}=k_{\alpha}/k$. $\overline{\varepsilon
}=\widehat{k}_{\alpha}\varepsilon_{\alpha\beta}\widehat{k}_{\beta}$ is the
reduced dielectric function. Eq.(\ref{epi}) shows that: (a) all three acoustic
branches contribute to $H_{\text{e-pie}}$; (b) $V_{\mathbf{k}s}^{\text{pie}%
}\varpropto k^{-1/2}$. The dielectric property of MAPbBr$_{3}$ is not very
anisotropic, we may approximate $\overline{\varepsilon}$ with Eqs.(\ref{di1},
\ref{di2}). Noticing Eqs.(\ref{e1}, \ref{di1}, \ref{di2}), we can see that
each term in $H_{\text{e-pie}}$ is proportional to $[1+2A(T-T_{c}%
)\varepsilon_{\infty}]^{-1}\thickapprox1$, a very weak monotonically
decreasing function of $T$.

For a non-piezoelectric material, the interaction $H_{\text{e-ph}}$ of a
carrier at position $\mathbf{x}$ with acoustic phonons is \cite{cal}%
\begin{equation}
H_{\text{e-Aph}}=\sum_{s\mathbf{k}}\frac{V_{s\mathbf{k}}^{\text{e-Aph}}}%
{\sqrt{\mathcal{N}}}(a_{s\mathbf{k}}e^{i\mathbf{k}\cdot\mathbf{x}%
}+a_{s\mathbf{k}}^{\dag}e^{i\mathbf{k}\cdot\mathbf{x}}),\label{ep}%
\end{equation}
where%
\begin{equation}
V_{s\mathbf{k}}^{\text{e-Aph}}=\sqrt{\frac{\hbar}{2c_{s}k}}\frac
{q^{2}z_{\kappa}k_{\gamma}e_{\kappa\gamma}^{s}(\mathbf{k})e^{i\mathbf{k}%
\cdot\mathbf{r}_{\kappa}}}{\varepsilon(c_{s}k,T)k^{2}\epsilon_{0}\Omega
\sqrt{M_{\kappa}}},\label{ep0}%
\end{equation}
$M_{\kappa}$ and $z_{\kappa}$ are the mass and effective nuclear charge of the
$\kappa$th atomic core in a primitive cell, $\mathbf{r}_{\kappa}$ is the
position vector of the $\kappa$th atom relative to the center-of-mass of a
primitive cell. $e_{\kappa\gamma}^{s}(\mathbf{k})$ is the polarization vector
of the $\kappa$th atom for $|s\mathbf{k}\rangle$ phonon. Eq.(\ref{ep}) shows
that (a) at least in the long wave limit, only the longitudinal acoustic (LA)
phonon contribute to $H_{\text{e-Aph}}$, later we will denote it as
$H_{\text{e-LA}}$; (b) Each term in $H_{\text{e-ph}}$ is proportional to
$k^{-3/2}$. Noticing Eqs.(\ref{di1}, \ref{di2}), we can see that each term in
$H_{\text{e-Aph}}$ is proportional to $2A(T-T_{c})$, a monotonically
increasing function of $T$.

\section{optical polarons as the main carriers}

\label{eop}

Assuming carriers are bare electrons (holes), a calculation shows that
piezoelectric scattering leads to $\mu\propto T^{-1.1}$ \cite{luy17}, which is
not fit the observed $\mu\propto T^{-1/2}$ in the piezoelectric tetragonal
phase of MAPbBr$_{3}$ \cite{yihal}. Thus, the main carriers in ABX$_{3}$
cannot be bare electrons (holes).

Let us show that in ABX$_{3}$ the electrons (holes) mainly exist as optical
polarons, neither piezoelectric polarons nor piezoelectric-optical polarons.
In any ionic crystal, piezoelectric or non-piezoelectric, the interaction
$H_{\text{e-LO}}$ of electron (hole) with the a branch of longitudinal optical
(LO) phonons is \cite{LLP}%
\begin{equation}
H_{\text{e-LO}}=\sum_{\mathbf{k}}\frac{V_{\mathbf{k}}^{\text{o}}}%
{\sqrt{\mathcal{N}}}(a_{\mathbf{k}\text{o}}e^{i\mathbf{k}\cdot\mathbf{x}%
}+a_{\mathbf{k}\text{o}}^{\dag}e^{i\mathbf{k}\cdot\mathbf{x}}),\label{op1}%
\end{equation}
where%
\begin{equation}
V_{\mathbf{k}}^{\text{o}}=iq[\frac{\hbar\omega_{\text{LO}}}{2\epsilon
_{0}\Omega}(\frac{1}{\varepsilon_{\infty}}-\frac{1}{\varepsilon_{0}}%
)]^{1/2}\frac{\mathbf{k}\cdot\mathbf{e}_{\mathbf{k}\text{o}}}{k^{2}%
}\label{op2}%
\end{equation}
is the coupling energy of electron (hole) with LO phonon in optical mode
$|$o$\mathbf{k}\rangle$, $\mathbf{k}$ is the wave vector of phonon,
$k=|\mathbf{k}|$. $\mathbf{e}_{\mathbf{k}\text{o}}$ is the polarization vector
of $|$o$\mathbf{k}\rangle$, $a_{\mathbf{k}\text{o}}$ is annihilation operator
of $|$o$\mathbf{k}\rangle$ phonon, $\omega_{\text{LO}}$ is the frequency of LO
phonon. $\varepsilon_{0}$ is the static dielectric constant, $q$ is the charge
of carrier, $\Omega$ is the volume of a primitive cell. $\mathcal{N}$ is the
number of primitive cells in a sample with volume $\mathcal{V}=\mathcal{N}%
\Omega$. $H_{\text{e-LO}}$ is not sensitive to use cubic or tetragonal
lattice. Later we will use the primitive cubic lattice, then $\Omega=a_{x}%
^{3}$, where $a_{x}$ is the length of basis vector along x-direction. One can
see from Eq.(\ref{op2}) that $V_{\mathbf{k}}^{\text{o}}\varpropto k^{-1}$.

We estimate the piezoelectric coupling energy in Eq.(\ref{epi}). The bulk
modulus of MAPbBr$_{3}$ is 20GPa, speed of longitudinal sound is
$c_{s}\thicksim$2000m/s \cite{fag17}, $\Gamma\thickapprox31.4\times10^{-12}%
$C$\cdot$N$^{-1}$ \cite{liuf16}, $\varepsilon_{0}=70$, $\varepsilon_{\infty
}=6.5$ \cite{ono},  $a_{x}=6.3$\AA , $\hbar\omega_{\text{LO}}\thicksim15.3$meV
\cite{wri}. Denote $k_{b}=\pi/a_{x}$ as the wave vector at the boundary of
Brillouin zone. Then, $V_{\mathbf{k}s}^{\text{pie}}=(\hbar/2Mc_{s}k_{b}%
)^{1/2}q\overline{\Xi}/2\overline{\varepsilon}\epsilon_{0}\thicksim13.9$meV,
and $V_{\mathbf{k}}^{\text{o}}\thicksim55.7$meV. Since in tetragonal phase of
MAPbBr$_{3}$, $H_{\text{e-pie}}<H_{\text{e-LO}}$, piezoelectric polaron or
piezoelectric-optical polaron do not form \cite{whit72,vo73,lica74}. In both
tetragonal and cubic phase, $H_{\text{e-LO}}>$ $\hbar\omega_{\text{LO}}$,
$H_{\text{e-pie}}$ and $H_{\text{e-Aph}}$, then the electrons (holes) mainly
exist as optical polarons \cite{pro,fro14,zhu,bre,sen,ost17,mob1} in both phases.

\section{residual interactions of polaron with phonons}

\label{pp}

We enumerate various interactions of polaron with other elementary
excitations. After the polaron transformation, the 1-LO phonon interaction
(\ref{op1}) disappeared, and the residual interaction $H_{\text{P-LO}}$ of
polaron with LO phonons is \cite{vo73}:
\[
H_{\text{P-LO}}=\sum_{\mathbf{kk}^{\prime}}O_{-\mathbf{k};\mathbf{k}^{\prime}}%
\]%
\begin{equation}
(a_{\mathbf{k}^{\prime}\text{o}}^{\dag}+a_{-\mathbf{k}^{\prime}\text{o}%
})(a_{\mathbf{k}\text{o}}^{\dag}+a_{-\mathbf{k}\text{o}})+\cdots, \label{ro}%
\end{equation}
where%
\begin{equation}
O_{-\mathbf{k};\mathbf{k}^{\prime}}=V_{-\mathbf{k}}^{\text{o}}V_{\mathbf{k}%
^{\prime}}^{\text{o}}\sum_{n\neq0}\frac{J_{-\mathbf{k}\text{o}}^{0n}%
J_{\mathbf{k}^{\prime}\text{o}}^{n0}}{E_{n}-E_{0}}, \label{ro1}%
\end{equation}
$E_{n}$ and $\varphi_{n}$ are the eigen values and eigen wave functions of the
polaron Hamiltonian $H_{\text{P}}$, $J_{\mathbf{k}^{\prime}\text{o}}^{n0}=\int
d^{3}x$ $\varphi_{n}^{\ast}(\mathbf{x})e^{i\mathbf{k}^{\prime}\cdot\mathbf{x}%
}\varphi_{0}(\mathbf{x})$ is a parameter describing the internal motion of
carrier in the potential well induced by phonons, the dependence on
$\mathbf{k}^{\prime}$ is weak \cite{mel73,vo73}.

In piezoelectric tetragonal phase, after the formation of optical polaron,
$H_{\text{e-pie}}$ becomes the piezoelectric interaction $H_{\text{P-pie}}$ of
polaron with acoustic phonons:%
\begin{equation}
H_{\text{P-pie}}=\sum_{s\mathbf{k}}\frac{V_{\mathbf{k}s}^{\text{P-pie}}}%
{\sqrt{\mathcal{N}}}(a_{s\mathbf{k}}e^{i\mathbf{k}\cdot\mathbf{x}%
}+a_{s\mathbf{k}}^{\dag}e^{i\mathbf{k}\cdot\mathbf{x}}),\label{ie1}%
\end{equation}
where%
\begin{equation}
V_{\mathbf{k}s}^{\text{P-pie}}=\frac{q\overline{\Xi}_{\gamma}e_{\gamma}%
^{s}(\mathbf{k})}{2\overline{\varepsilon}\epsilon_{0}}\sqrt{\frac{\hbar
}{2Mc_{s}k}}.\label{ie2}%
\end{equation}
Because the charge of electronic (hole) polaron is same as the charge of an
electron (hole), Eqs.(\ref{ie1},\ref{ie2}) have the same form as
Eqs.(\ref{epi},\ref{epi1}). Eqs.(\ref{op2},\ref{ro1},\ref{ie2}) indicates that
$H_{\text{P-pie}}$ is stronger than $H_{\text{P-LO}}$.

In non-piezoelectric cubic phase, after the formation of polaron,
$H_{\text{e-LA}}$ becomes the poarlon-LA phonon interaction $H_{\text{P-LA}}$:%
\begin{equation}
H_{\text{P-LA}}=\sum_{\mathbf{k}}\frac{V_{\mathbf{k}}^{\text{LA}}}%
{\sqrt{\mathcal{N}}}(a_{\mathbf{k}\text{LA}}e^{i\mathbf{k}\cdot\mathbf{x}%
}+a_{\mathbf{k}\text{LA}}^{\dag}e^{i\mathbf{k}\cdot\mathbf{x}}), \label{PLA}%
\end{equation}
where%
\begin{equation}
V_{\mathbf{k}}^{\text{LA}}=\sqrt{\frac{\hbar}{2c_{l}k}}\frac{q^{2}z_{\kappa
}k_{\gamma}e_{\kappa\gamma}^{s}(\mathbf{k})e^{i\mathbf{k}\cdot\mathbf{r}%
_{\kappa}}}{\varepsilon(c_{l}k,T)k^{2}\epsilon_{0}\Omega\sqrt{M_{\kappa}}}.
\label{PLA1}%
\end{equation}
Eqs.(\ref{PLA},\ref{PLA1}) have the same form as Eqs.(\ref{ep},\ref{ep0}).
According to Eq.(\ref{di2}), the screening becomes weaker at higher
temperature. Thus, in cubic phase, $H_{\text{P-LA}}$ is comparable to
$H_{\text{P-LO}}$.

By means of golden rule, one can show that for $n_{e}<10^{23}$cm$^{-3}$ and
moderate defect concentration ($\leq10^{20}$cm$^{-3}$), the scattering
probabilities per unit time produced by the Coulomb interaction between
polarons and by the interaction of polaron with defects are much smaller than
those caused by $H_{\text{P-LO}}$ and by the interaction of polaron with
acoustic phonons \cite{mob1}. For the aim of calculating mobility, we can
ignore the Coulomb interaction between polarons and the interaction of polaron
with defects \cite{zhu,bre}.

\section{mobility in piezoelectric phase}

\label{mpie}

If a polaron does not interact with other objects, its state can be
characterized by wave vector $\mathbf{p}$ \cite{v10}. Denote $f_{\mathbf{p}}$
as the distribution function of polaron. Each interaction of polaron produces
a change in distribution function $f_{\mathbf{p}}$, corresponds to an
collision integral in Boltzmann equation \cite{v10}. If there are $N$ types of
interactions of polaron, the total change rate $\nu(T)$ of $f_{\mathbf{p}}$ is
$\nu(T)=\sum_{j=1}^{N}\nu_{j}(T)$, where $\nu_{j}(T)$ is the change rate of
$f_{\mathbf{p}}$ caused by the $j$th interaction. The mobility $\mu(T)$ of
polaron is determined by: $\mu(T)=q/m_{\text{P}}\nu(T)$, where $q$ and
$m_{\text{P}}$ are the charge and effective mass of polaron \cite{LLP,v10}.

In tetragonal piezoelectric phase, the charge transport of polaron is
controlled by $H_{\text{P-pie}}$. The rate $\nu_{\text{P-pie}}=(\partial
f_{\mathbf{p}}/\partial t)_{\text{P-pie}}$ of change in the distribution
function $f_{\mathbf{p}}$ of polaron with wave vector $\mathbf{p}$ caused by
$H_{\text{P-pie}}$ is \cite{v10}%

\begin{equation}
(\frac{\partial f_{\mathbf{p}}}{\partial t})_{\text{P-pie}}=-\sum
_{s\mathbf{k}}\frac{\partial N_{0}(\hbar c_{s}k)}{\partial\hbar c_{s}k}%
[f_{0}(\mathbf{p}^{\prime})-f_{0}(\mathbf{p})]\label{ep1}%
\end{equation}%
\[
\{w(\mathbf{p}^{\prime},\mathbf{k};\mathbf{p})(\varphi_{\mathbf{p}^{\prime}%
}-\varphi_{\mathbf{p}}+\chi_{s\mathbf{k}})\delta(E_{\mathbf{p}}-E_{\mathbf{p}%
^{\prime}}-\hbar c_{s}k)
\]%
\[
-w(\mathbf{p}^{\prime};\mathbf{p},\mathbf{k})(\varphi_{\mathbf{p}^{\prime}%
}-\varphi_{\mathbf{p}}-\chi_{s\mathbf{k}})\delta(E_{\mathbf{p}}-E_{\mathbf{p}%
^{\prime}}+\hbar c_{s}k)\}
\]
where $f_{0}$ and $N_{0}$ are the equilibrium distribution functions at
temperature $T$ for polaron and phonons, $E_{\mathbf{p}}=$ $\hbar^{2}%
p^{2}/2m_{\text{P}}$ is the kinetic energy of polaron, $m_{\text{P}}$ is the
mass of polaron. $\varphi$ and $\chi$ describe the deviations of
$f_{\mathbf{p}}$ and $N_{s\mathbf{k}}$ from equilibrium%
\begin{equation}
f_{\mathbf{p}}-f_{0}(\varepsilon)=-\frac{\partial f_{0}(E_{\mathbf{p}}%
)}{\partial E_{\mathbf{p}}}\varphi_{\mathbf{p}},\label{df}%
\end{equation}
and%
\begin{equation}
N_{s\mathbf{k}}-N_{0}(s\mathbf{k})=-\frac{\partial N_{0}(\hbar c_{s}%
k)}{\partial\hbar c_{s}k}\chi_{s\mathbf{k}}.\label{dn}%
\end{equation}
At temperature $T$, the characteristic energies $\varphi_{\mathbf{p}}$ and
$\chi_{s\mathbf{k}}$ are order of $k_{B}T$, then $\varphi_{\mathbf{p}^{\prime
}}-\varphi_{\mathbf{p}}+\chi_{s\mathbf{k}}\thicksim k_{B}T$ \cite{v10,mob1}.
Because the density $n_{e}$ of polarons gas is rare, polaron momentum is much
smaller than $k_{b}=\pi/a_{x}$. One can neglect reciprocal processes.
$\mathbf{p}^{\prime}=\mathbf{p}+\mathbf{k}$ is implied in Eq.(\ref{ep1}), then%
\begin{equation}
\delta(E_{\mathbf{p}}-E_{\mathbf{p}^{\prime}}+\hbar c_{s}k)\label{de}%
\end{equation}%
\[
=\frac{\delta(k)+\delta\lbrack k-(\frac{2m_{\text{P}}c_{s}}{\hbar}%
-2p\cos\theta)]}{|\frac{\hbar^{2}p\cos\theta}{m_{\text{P}}}-\hbar c_{s}|}.
\]
The probability coefficient $w(\mathbf{p}^{\prime},\mathbf{k};\mathbf{p})$ is
defined by%
\begin{equation}
\frac{2\pi}{\hbar}|\langle\mathbf{p}^{\prime},s\mathbf{k}|H_{\text{P-pie}%
}|\mathbf{p}\rangle|^{2}=w(\mathbf{p}^{\prime},s\mathbf{k};\mathbf{p}%
)(N_{s\mathbf{k}}+1),\label{w1}%
\end{equation}
or%
\begin{equation}
\frac{2\pi}{\hbar}|\langle\mathbf{p}^{\prime},s\mathbf{k}|H_{\text{P-pie}%
}|\mathbf{p}\rangle|^{2}=w(\mathbf{p}^{\prime},s\mathbf{k};\mathbf{p}%
)N_{s\mathbf{k}},\label{w2}%
\end{equation}
the occupation number and energy consideration delta function are factored out
from the transition probability. Using Eqs.(\ref{ie2},\ref{w1}), one has%
\begin{equation}
w\thicksim\frac{2\pi}{\hbar}\frac{\hbar}{2\mathcal{N}Mc_{s}k}[\frac
{q\overline{\Xi}_{\gamma}e_{\gamma}^{s}(\mathbf{k})}{2\overline{\varepsilon
}\epsilon_{0}}]^{2}\label{w}%
\end{equation}%
\[
\thicksim\frac{2\pi}{\hbar}\frac{\hbar}{2\mathcal{N}Mc_{s}k}(\frac{qc\Lambda
}{2})^{2}.
\]
Take the direction of wave vector $\mathbf{p}$ of the initial state of polaron
as the polar axis, then $\sum_{\mathbf{k}}$ $\rightarrow(2\pi)^{-3}%
\mathcal{V}k^{2}dkd\cos\theta d\phi$, where ($\theta,\phi$) are the polar and
azimuth angels of $\mathbf{k}$ relative to $\mathbf{p}$. In halide
perovskites, even $\hbar c_{s}k_{b}=75$K$\ll k_{B}T$ in experimental
temperature range 160-280K. Since $k\leq k_{b}$, then for all $\mathbf{k}$,%
\begin{equation}
\frac{\partial N_{0}(\hbar c_{s}k)}{\partial\hbar c_{s}k}\thicksim\frac
{k_{B}T}{(\hbar c_{s}k)^{2}}.\label{bs}%
\end{equation}

For the thermal generated \textit{intrinsic} carriers, the density $n_{e}$ of
electrons (i.e. polarons) is temperature dependent \cite{ash}. Because $n_{e}$
is small, the polaron gas is non-degenerate, the equilibrium distribution
$f_{0\mathbf{p}}$ of polaron with wave vector $\mathbf{p}$ is \cite{ash}%
\begin{equation}
f_{0\mathbf{p}}\thickapprox\exp(-\hbar^{2}p^{2}/2m_{\mathbf{P}}k_{B}T),
\label{ind}%
\end{equation}
where $m_{\mathbf{P}}$ is the mass of polaron. The delta function in
Eq.(\ref{ep1}) requires $E_{\mathbf{p}^{\prime}}=E_{\mathbf{p}}\pm\hbar
c_{s}k$, then
\[
f_{0}(\mathbf{p}^{\prime})-f_{0}(\mathbf{p})=\frac{\partial f_{0}%
(E_{\mathbf{p}})}{\partial E_{\mathbf{p}}}(E_{\mathbf{p}^{\prime}%
}-E_{\mathbf{p}})
\]%
\begin{equation}
\thicksim\mp\exp(-\frac{\hbar^{2}p^{2}}{2m_{\text{P}}k_{B}T})\frac{\hbar
c_{s}k}{k_{B}T}. \label{fer}%
\end{equation}
Substitute Eqs.(\ref{df},\ref{dn},\ref{de},\ref{w},\ref{bs},\ref{fer}) into
Eq.(\ref{ep1}), and carry out the momentum integral, one obtains
\begin{equation}
\nu_{\text{P-pie}}=\frac{e^{-3/2}\sqrt{3m_{\text{P}}k_{B}T}}{32\pi\rho}
\label{fre}%
\end{equation}%
\[
\sum_{s}(\frac{qc\Lambda}{\hbar c_{s}})^{2}\ln\left\vert \frac{\sqrt
{3m_{\text{P}}k_{B}T}-m_{\text{P}}c_{s}}{\sqrt{3m_{\text{P}}k_{B}%
T}+m_{\text{P}}c_{s}}\right\vert ,
\]
where $e$ is the base of natural logarithm, $\rho=M/\Omega$ is the density of
halide perovskites. In derive Eq.(\ref{fre}), energy equipartition theorem has
been used: the average momentum $\hbar p$ of polaron is taken as $\hbar
p=\sqrt{3m_{\text{P}}k_{B}T}$. The logarithm factor in Eq.(\ref{fre}) only
weakly depends on temperature, therefore $\mu\propto T^{-1/2}$ in
piezoelectric tetragonal phase. One can see that two critical ingredients for
$\mu\propto T^{-1/2}$ are (1) $V_{\mathbf{k}s}^{\text{P-pie}}\propto k^{-1/2}%
$; and (2) the distribution function for \textit{intrinsic} carriers is
Eq.(\ref{ind}). The piezoelectric electron-acoustic phonon interaction has
been discussed with deformation potential approximation in which
$H_{\text{e-pie}}\propto k^{-1}$ \cite{hus61} in contrast to Eq.(\ref{epi1}).
The square of transition element $|M(\mathbf{k},\mathbf{k}^{\prime}%
)|^{2}=E_{1}^{2}k_{B}T/2c$ was used to derive $\mu\propto T^{-1/2}$, where $c$
is elastic constant, $E_{1}$ is the deformation potential energy \cite{hus61}.
The key assumption behind this derivation is that the polarization-strain
tensor does not depend on temperature in contrast to Eq.(\ref{e1}). A future
knowledge of $\Xi(T)$ in halide perovskite could resolve which reasoning is
more plausible.

\begin{figure}
[ht]\centering
\subfigure[]{\includegraphics[width=0.23\textwidth]{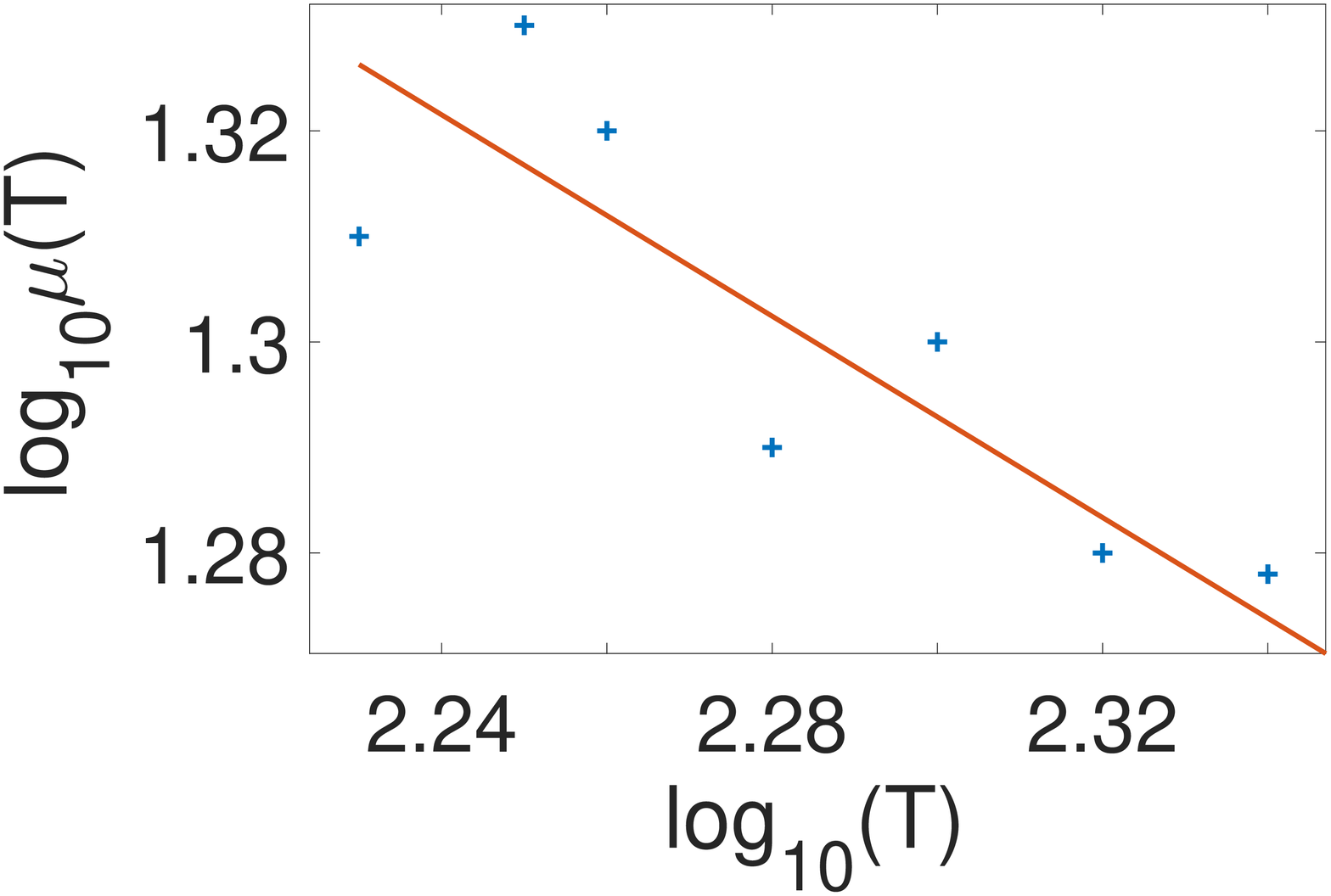}\label{tem}}
\subfigure[]{\includegraphics[width=0.23\textwidth]{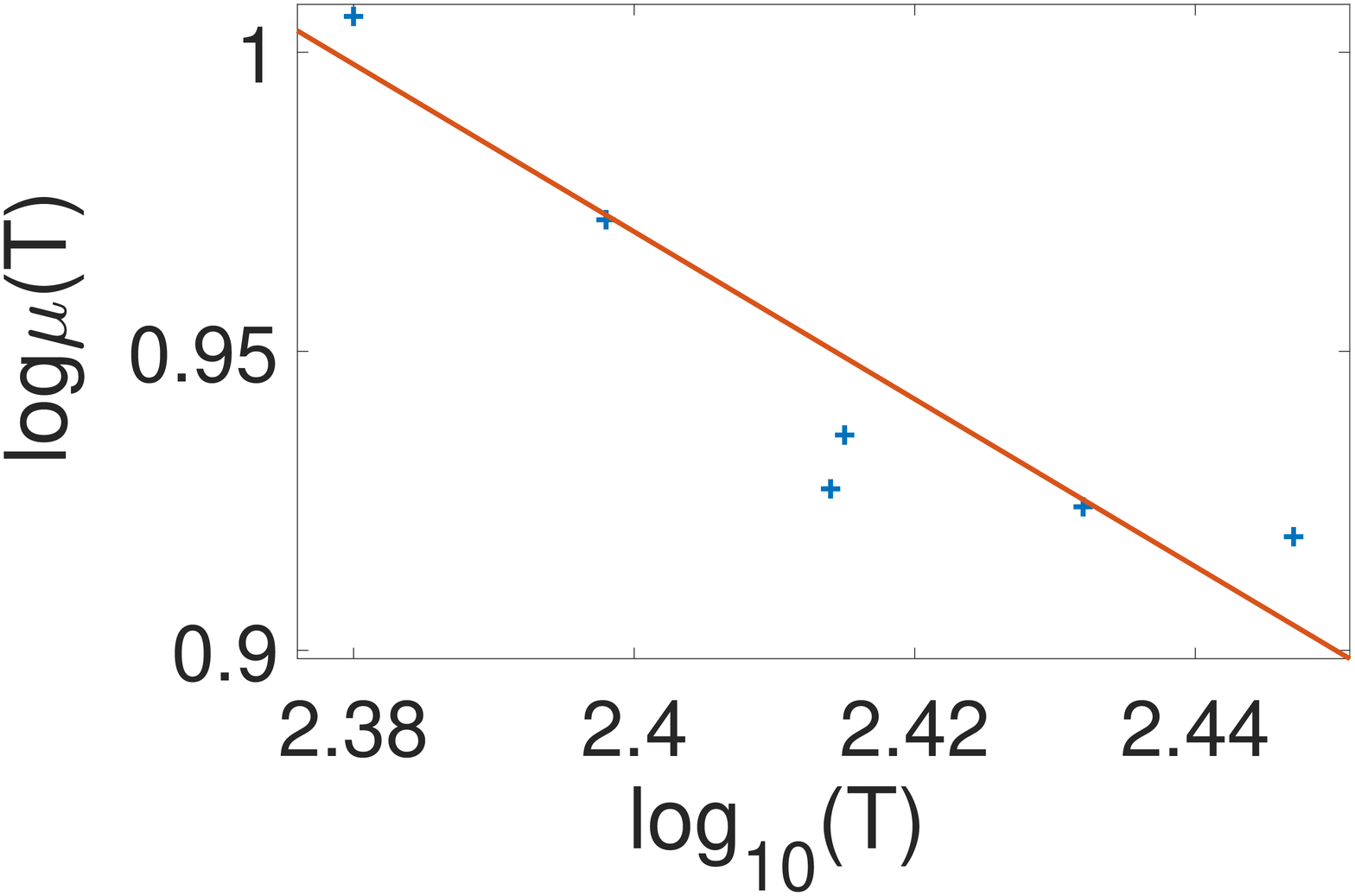}\label{cum}}
\caption{Mobility $\mu$ vs. temperature $T$ of MAPbBr$_{3}$, experimental data (+) taken from \cite{yihal}. (a) piezoelectric tetragonal phase, solid line is a fitting based on Eq.(\ref{fre}); (b) non-piezoelectric cubic phase, solid line is a fitting based on Eqs.(\ref{fla},\ref{flo}). }
\label{pimo}
\end{figure}

In Fig.\ref{tem}, we compare the observed mobility for intrinsic carriers in
tetragonal phase of MAPbBr$_{3}$ with that estimated from Eq.(\ref{fre}). Due
to lack of components of tensor $\Xi_{\gamma\alpha\beta}$, in the calculation,
elastic modulus $c$ taken as 20GPa \cite{fag17}, $\Gamma\thickapprox
31.4\times10^{-12}$C$\cdot$N$^{-1}$ \cite{liuf16}, $\Xi=6.28\times10^{-1}%
$C$\cdot$m$^{-2}$, $m_{\text{P}}=3.8m$, $c_{s}=2000$m/s are used, where $m$ is
mass of electron. The agreement is reasonable well. For the \textit{intrinsic}
carriers in MAPbI$_{3}$, $\mu\propto T^{-0.42}$ has been observed at
$50<T<150$K \cite{laov}, it seems indicate that the orthorhombic phase is also piezoelectric.

\section{mobility in non-piezoelectric phase}

\label{mni}

In Sec.\ref{pp}, we noticed that $H_{\text{P-LO}}$ is comparable to
$H_{\text{P-LA}}$, both contribute to $\partial f_{\mathbf{p}}/\partial t$
significantly. By similar procedure deriving Eq.(\ref{fre}), the change rate
$\nu_{\text{P-LA}}=(\partial f_{\mathbf{p}}/\partial t)_{\text{P-LA}}$ of
distribution function $f_{\mathbf{p}}$ of polaron caused by $H_{\text{P-LA}}$
can be obtained:%
\begin{equation}
\nu_{\text{P-LA}}\thicksim\frac{e^{-3/2}k_{B}T^{3}A^{2}m_{\text{P}}^{2}}{2\pi
c_{l}[3m_{\text{P}}k_{B}T-(m_{\text{P}}c_{l})^{2}]^{2}\Omega}\label{fla}%
\end{equation}%
\[
\langle\lbrack\frac{q^{2}z_{\kappa}e^{i\mathbf{k}\cdot\mathbf{r}_{\kappa}}%
}{\epsilon_{0}\sqrt{M_{\kappa}}}]^{2}\rangle_{\text{av}},
\]
where an average over all $\mathbf{k}$ points is taken in the last factor. If
$H_{\text{P-LA}}$ was the solely interaction changing $f_{\mathbf{p}}$,
Eq.(\ref{fla}) would imply $\mu(T)=q/m_{\text{P}}\nu(T)\propto T^{-1}$.

The change rate $\nu_{\text{P-LO}}=(\partial f_{\mathbf{p}}/\partial
t)_{\text{P-LO}}$ of distribution function caused by $H_{\text{P-LO}}$
involves 2-LO phonons:%
\[
(\frac{\partial f_{\mathbf{p}}}{\partial t})_{\text{P-LO}}=\sum_{\mathbf{kk}%
^{\prime}}w(\mathbf{p}^{\prime}\mathbf{k}^{\prime};\mathbf{pk})\delta
(\epsilon_{\mathbf{p}^{\prime}}+\hbar\omega_{\mathbf{k}^{\prime}}%
-\epsilon_{\mathbf{p}}-\hbar\omega_{\mathbf{k}})
\]%
\[
n_{0\mathbf{p}}(1-n_{0\mathbf{p}^{\prime}})N_{0\mathbf{k}}(1+N_{0\mathbf{k}%
^{\prime}})\frac{\varphi_{\mathbf{p}^{\prime}}+\chi_{\mathbf{k}^{\prime}}%
-\chi_{\mathbf{k}}-\varphi_{\mathbf{p}}}{k_{B}T}%
\]%
\begin{equation}
+\sum_{\mathbf{kk}^{\prime}}w(\mathbf{p}^{\prime},\mathbf{k},\mathbf{k}%
^{\prime};\mathbf{p})\delta(\epsilon_{\mathbf{p}^{\prime}}+\hbar
\omega_{\mathbf{k}^{\prime}}+\hbar\omega_{\mathbf{k}}-\epsilon_{\mathbf{p}})
\label{lo}%
\end{equation}%
\[
n_{0\mathbf{p}}(1-n_{0\mathbf{p}^{\prime}})(1+N_{0\mathbf{k}}%
)(1+N_{0\mathbf{k}^{\prime}})\frac{\varphi_{\mathbf{p}^{\prime}}%
+\chi_{\mathbf{k}^{\prime}}+\chi_{\mathbf{k}}-\varphi_{\mathbf{p}}}{k_{B}T},
\]
the first term is caused by scattering of LO-phonon by polaron $\mathbf{p}%
+\mathbf{k\rightleftarrows p}^{\prime}+\mathbf{k}^{\prime}$, where
$\mathbf{p}$ and$\ \mathbf{p}^{\prime}$ are wave vectors of polaron before and
after scattering, $\mathbf{k}$ and$\ \mathbf{k}^{\prime}$ are wave vectors of
the incident and outgoing LO phonons. The 2nd term is caused by the emitting
and absorbing 2-LO phonons $\mathbf{p\rightleftarrows p}^{\prime}%
+\mathbf{k}^{\prime}\mathbf{+\mathbf{k}}$. Noticing the dispersion of LO
phonon is negligible, by a similar procedure deriving Eq.(\ref{fre}), one
finds
\begin{equation}
\nu_{\text{P-LO}}\thicksim\frac{e^{-3/2}(k_{B}T)^{3/2}k_{b}}{(2\pi)^{3}}%
\sqrt{\frac{m_{\text{P}}}{3}} \label{flo}%
\end{equation}%
\[
\lbrack\frac{q^{2}}{2\epsilon_{0}\hbar}(\frac{1}{\varepsilon_{\infty}}%
-\frac{1}{\varepsilon_{0}})]^{2}|\sum_{n\neq0}\frac{J_{-\mathbf{k}\text{o}%
}^{0n}J_{\mathbf{k}^{\prime}\text{o}}^{n0}}{E_{n}-E_{0}}\mid^{2},
\]
where $k_{b}=\pi/a_{x}$ is the wave vector at boundary of Brillouin zone. If
$H_{\text{P-LO}}$ was the solely interaction changing $f_{\mathbf{p}}$,
Eq.(\ref{flo}) would imply $\mu(T)=q/m_{\text{P}}\nu(T)\propto T^{-3/2}$. But
the total changing rate $\nu(T)$ of $f_{\mathbf{p}}$ is a sum of those caused
by $H_{\text{P-LO}}$ and by $H_{\text{P-LA}}$: $\nu(T)=\nu_{\text{P-LO}}%
+\nu_{\text{P-LA}}$. Then $\mu(T)\propto T^{-\gamma}$, $\gamma$ is a number
between 1 and 3/2. For the \textit{intrinsic} carriers in non-piezoelectric
cubic phase of MAPbBr$_{3}$, $\gamma=1.4$ \cite{yihal} indicates that
$H_{\text{P-LO}}$ seems more important than $H_{\text{P-LA}}$. $\gamma=1.18$
has been observed for \textit{intrinsic} carriers in orthorhombic phase
MAPbI$_{3}($Cl$)$ below 150 K \cite{kar}. As noticed earlier, piezoelectric
effect does not appear in the orthorgonal phase due to the rigid structure.
Then $f_{\mathbf{p}}$ is changed by $H_{\text{P-LA}}$ and $H_{\text{P-LO}}$
rather than $H_{\text{P-pie}}$. $\gamma=1.18$ indicates that $H_{\text{P-LA}}$
is larger than $H_{\text{P-LO}}$.

The piezoelectric polaron-acoustic phonon interaction (\ref{ie2}) has weaker
temperature dependence than the ordinary polaron-LA phonon interaction
(\ref{PLA1}), which is one of key factors leading to weaker temperature
dependence of mobility in tetragonal phase. In Fig.\ref{cum}, we compare the
observed mobility \cite{yihal} with that expected from Eqs.(\ref{fla}%
,\ref{flo}). Except the parameters used in Sec.\ref{mpie}, $z_{\kappa}=2$,
$M_{\kappa}$ taken as the mass of Pb. $O_{-\mathbf{k};\mathbf{k}^{\prime}%
}\thickapprox10$meV is used due to lack of knowledge of $\varphi_{n}$ to
calculate $J_{\mathbf{k}^{\prime}\text{o}}^{n0}$. The agreement is reasonable well.

\section{\textit{intrinsic} carriers vs. photo-generated carriers}

When a halide perovsikte is not exposed to light, the properties of material
are different to those of material which is shined by light in three aspects.
When a material is kept in dark, the concentration of \textit{intrinsic}
carriers is determined by temperature \cite{ash,yihal}, its distribution
function is give by Eq.(\ref{ind}). The relatively lower concentration of
intrinsic carriers (10$^{9}$-10$^{12}$cm$^{-3}$) \cite{yihal} has two
consequences: (i) piezoelectric effect exists in tetragonal phase but not in
cubic phase. The $T$ and $\mathbf{k}$ dependence of the piezoelectric
polaron-acoustic interaction (\ref{ie1}) are different to those of ordinary
polaron-LA phonon interaction (\ref{PLA}); and (ii) the density of states of
polarons is lower, the energy conservation delta functions in collision
integral is difficult to satisfy. To remove delta function, one has to carry
out $k\mathbf{-}$integral precisely.

When a sample is illuminated by a beam of light, the majority of carriers are
photo-generated, the density $n_{e}$ of electrons is fixed by the incident
flux of photons, not depend on temperature. In normal operation condition,
$n_{e}<$10$^{18}$cm$^{-3}$ \cite{oga}, the polaron gas is non-degenerate. The
equilibrium distribution function is \cite{mob1}
\begin{equation}
f_{0\mathbf{p}}\thickapprox n_{e}4\pi^{3/2}\hbar^{3}\frac{\exp(-\hbar^{2}%
p^{2}/2m_{\mathbf{P}}k_{B}T)}{(2m_{\mathbf{P}}k_{B}T)^{3/2}}. \label{phd}%
\end{equation}
The relatively higher concentration (10$^{13}$-10$^{17}$cm$^{-3}$) of
photo-generated carriers \cite{oga,mil,sav} has two consequences: (i) even in
the tetragonal phase lacking of inversion center, piezoelectric effect is
suppressed by the mobile charges. In both tetragonal and cubic phases, the
polaron-acoustic phonon interaction is given by Eq.(\ref{PLA}); and (ii) the
density of states of polaron is higher, it is legitimate to replace the delta
function in $H_{\text{P-LA}}-$collision integral with $(\hbar c_{l}k_{b}%
)^{-1}$ \cite{mob1}. When we execute Boltzmann equation analysis, three
different aspects between \textit{intrinsic} carriers and photo-generated
carriers naturally produce their different $\mu(T)$ behaviors.

In summary, for the \textit{intrinsic} carriers in MAPbBr$_{3}$ we resolved
the puzzle that $\mu(T)\propto T^{-1/2}$ in tetragonal phase and
$\mu(T)\propto T^{-1.4}$ in cubic phase. The current scheme is consistent with
a previous research on the mobility of photo-generated carriers \cite{mob1},
further clarifies the mechanism of charge transport in halide perovsikte, and
will help improve the charge separation efficiency in perovskite based solar cells.

$^{\dagger}$ zmh1999238@163.com

\bibliographystyle{apsrev4-1}
\bibliography{refpie}

\end{document}